\documentclass[epj,nopacs]{svjour}
\usepackage{graphicx}
\begin{document}
\title{The scar mechanism revisited}
\author{F. Borondo\inst{1}\fnmsep\thanks{\email{f.borondo@uam.es}}
\and
D. A. Wisniacki\inst{2}\fnmsep\thanks{\email{wisniacki@df.uba.ar}}
\and
E. G. Vergini\inst{3}\fnmsep\thanks{\email{eduardogerman.vergini@upm.es}}
\and
R. M. Benito\inst{3}\fnmsep\thanks{\email{rosamaria.benito@upm.es}}
}
\institute{Departamento de Qu\'{\i}mica,
Facultad de Ciencias C--IX,
Universidad Aut\'onoma de Madrid, CANTOBLANCO, 28049 Madrid, Spain
\and
Departamento de F\'\i sica ``J. J. Giambiagi'', FCEN,
UBA, Pabell\'on 1, Ciudad Universitaria, 1428 Buenos Aires, Argentina
\and
Grupo de Sistemas Complejos
 and Departamento de F\'{\i}sica,
 Escuela T\'ecnica Superior de Ingenieros
 Agr\'onomos, Universidad Polit\'ecnica de Madrid,
 28040 Madrid, Spain}
\abstract{
Unstable periodic orbits are known to originate scars on some eigenfunctions
of classically chaotic systems through recurrences causing that some part
of an initial distribution of quantum probability in its vicinity returns
periodically close to the initial point.
In the energy domain, these recurrences are seen to accumulate quantum
density along the orbit by a constructive interference mechanism when the
appropriate quantization (on the action of the scarring orbit) is fulfilled.
Other quantized phase space circuits, such as those defined by homoclinic
tori, are also important in the coherent transport of quantum density
in chaotic systems.
The relationship of this secondary quantum transport mechanism with the
standard mechanism for scarring is here discussed and analyzed.
}
\maketitle
\section{Introduction}
  \label{sec:intro}
The quantum calculation performed on the Bunimovitch stadium billiard
\cite{bun} by McDonald and Kaufmann \cite{McDK} showed, quite  surprisingly,
that some of the eigenfunctions present anomalously high density along
certain (unstable) periodic orbits (PO) of the system.
This example demonstrates in a very dramatic way the existence of a
correspondence between quantum and classical mechanics \cite{Gutzwiller}.

Five years later, Heller \cite{Heller1} established the basis for the
corresponding theory \cite{Heller2} using semiclassical arguments,
and coined the term ``scarring'' to name this phenomenon.
In this seminal work, the physical mechanism for this enhanced localization
was explained, by considering the dynamical evolution of a
localized wave packet, $|\phi(0) \rangle$, initially launched in the
vicinity of the scarring PO.
The center of such packet follows for some time the classical path dictated
by the PO, so that part of the initial quantum density distribution
periodically returns next to the initial region.
This fact originates recurrences in the self--correlation function
$C(t)=\langle \phi(0)|\phi(t) \rangle$, whose intensity decreases
exponentially with time due to the competing mechanism induced by the
unstable dynamics, characterized by the Lyapunov exponent $\lambda$,
in the vicinity of the scarring PO.
This discussion can be made more quantitative considering the relationship
between the two associated time scales, so that it can be stated that
for a scar to be seen a necessary condition is given by the criterion
$\lambda/\omega << 1$, where $\omega=2\pi/\tau$ is the frequency
associated to the PO.

Notice that the above arguments were formulated in the time domain,
but they can be translated into the energy domain by considering
the spectrum associated to $C(t)$
%
\begin{equation}
  I(E)= \frac{1}{2\pi} \; \int \; dt \; e^{iEt/\hbar} \; C(t).
  \label{eq:1}
\end{equation}
This spectrum appears resolved in narrow peaks of width
proportional to $\hbar\lambda$, occurring at intervals of $\hbar\omega$,
as pictorially depicted in Fig.~22 of Ref.~\cite{Heller2}.
Moreover, when this low resolution spectrum is projected in the system
eigenspectrum \cite{Polavieja}, the eigenstate intensities under these
bands are enhanced by a factor of $\omega/\lambda$ with respect to a
random distribution \cite{Kaplan}.

One important issue of this alternative view in the energy domain is that it
shows very clearly the relevance of the quantization of the action, $S(E)$,
along the scarring PO \cite{Gutzwiller} in our problem.
The energies, $E_{\rm BS}$, for which this quantity fulfills the
Bohr--Sommerfeld (BS) quantization condition
%
\begin{equation}
  S(E_{\rm BS})= 2 \pi \hbar \left(n+\frac{\nu}{4} \right),
  \label{eq:2}
\end{equation}
being $n$ the corresponding quantum number and $\nu$ the Maslov index
associated to the orbit, mark the position of the centers of the low
resolution bands appearing in (\ref{eq:1}).
The reason for this is clear, since condition (\ref{eq:2}) guarantees the
coherence of the probability recurrences along the PO, which is responsible
for a constructive interference of the returning wave, causing the
building up of probability density constituting the scar.
This way of proceeding is similar to the way in which the optical problems,
such as interference, are treated.
There, the notion of optical path, analogous to our action, followed by
the different waves all propagating at the same energy, is used.
In this way, any reference to time can be avoided.

Another important contribution to the theory of scars is due to Bogomolny
\cite{Bogomolny}, who derived an explicit expression for the PO
contributions to the smoothed quantum probability
density over small ranges of space and energy
(i.e.\ average over a large number of eigenfunctions).
A corresponding theory for Wigner functions was developed by Berry
\cite{Berry}.
Something interesting to remark about this work in connection with the present
paper is that its author noticed the existence of fringes as one moves away
from the scarring PO, either in position or energy.
Also, Prado and Keating focussed recently on the influence of bifurcations
on scarring \cite{Keating}, something of outmost relevance for (realistic)
mixed systems.

On the experimental side, many practical applications in areas of much
interest have been described in the literature concerning scar theory.
Among them, nanodevices \cite{RTD}, optical microcavities \cite{lasers},
or optical fibers \cite{Doya} deserve a special mention.

As discussed above, reference \cite{Heller1} mainly focussed on the
probability density recurrently returning next to the initial point at
time intervals corresponding to the scarring PO period,
paying no attention to the fate of the remaining density
leaving the PO dynamically unstable region.
Obviously, this density can be thought as formed by different pieces,
each of which would ramble through the available phase space by a
different pathway, each one of them with its own characteristic complexity.
Actually, by summing up all these contributions Heller and Tomsovic
\cite{Tomsovic} were able to construct a semiclassical approximation
to wave functions valid past the limit of the Ehrenfest time \cite{tE},
something that came as a surprise since at that point the underlying chaos
has had time to develop structure at a scale much finer than $\hbar$.

The purpose of this paper is to study what happens with the portion of the
quantum probability density first expelled from the vicinity of the scarring
PO, but returning back to the initial region by the shortest circuit
that is different from that defined by the scarring PO itself.
In this sense, our goal is to present a complementary view to that
provided by the standard scarring mechanism.
The phenomenon that will be described constitutes a second order effect,
since the amount of quantum density involved is orders of magnitude smaller
than that transported along the scarring PO.
The key point is to consider the problem in the energy domain, so that
the importance of homoclinic excursions related to the scarring PO
(with the associated propagation of quantum waves) can be quantitatively
analyzed, and the relevance of the quantization of these phase space circuits
gauged.

The organization of this paper is as follows.
In Sect.~\ref{sec:model} we briefly described the model used in our
calculations.
In Sect.~\ref{sec:classical} we discuss some aspects of the classical
dynamics of this system which are relevant to our work.
We continue with Sect.~\ref{sec:quantum}, in which we present and analyzed
our quantum results.
We summarize our main conclusions in Sect.~\ref{sec:conclusions}.
\section{Model}
   \label{sec:model}

\begin{figure}
   \begin{center}
   \resizebox{0.63\columnwidth}{!}{ \includegraphics{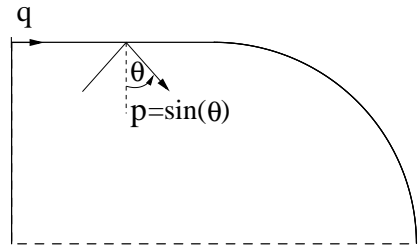} }
   \end{center}
 \caption{Desymmetrized Bunimovitch stadium billiard and definition of the
   Birkhoff coordinates $(q,p)$ used in our study.
   We focus our attention on the dynamics influenced by the horizontal
   periodic orbit drawn in dashed line.}
 \label{fig:1}
\end{figure}

The model used in our calculations consist of a particle of mass
$M=\hbar^2/2$ confined in a desymmetrized Bunimovitch stadium billiard,
defined by the radius of the circular part, $r=1$, and the enclosed area,
$1+\pi/4$, as shown in Fig.~\ref{fig:1}.
Note that in this units system the energy of the particle is simply
given by the square of the wave number, $E=k^2$.
This billiard constitutes a standard paradigm in the field of quantum chaos,
and it has been shown to display all characteristics (ergodic, mixing, \ldots)
of the chaotic hierarchy at classical level \cite{ergodic}.

To study the classical and quantum dynamics of this system,
Birkhoff coordinates (on the boundary) \cite{ergodic}, ($q,p$),
are customarily used.
They can be seen superimposed in Fig.~\ref{fig:1}.
Coordinate $q$, is the arc length coordinate, usually measured starting from
the upper left corner of the stadium,
and $p={\bf p}\cdot{\bf{\hat t}}/|{\bf p}|$ is the fraction of tangential
momentum.

\section{Classical dynamical aspects}
   \label{sec:classical}

In this paper, we will focus our attention on the dynamics influenced by the
horizontal PO, which coincides in our case with the axis for the $x$ coordinate.
See dashed line in Fig.~\ref{fig:1}.

\begin{figure}
   \begin{center}
   \resizebox{0.95\columnwidth}{!}{ \includegraphics{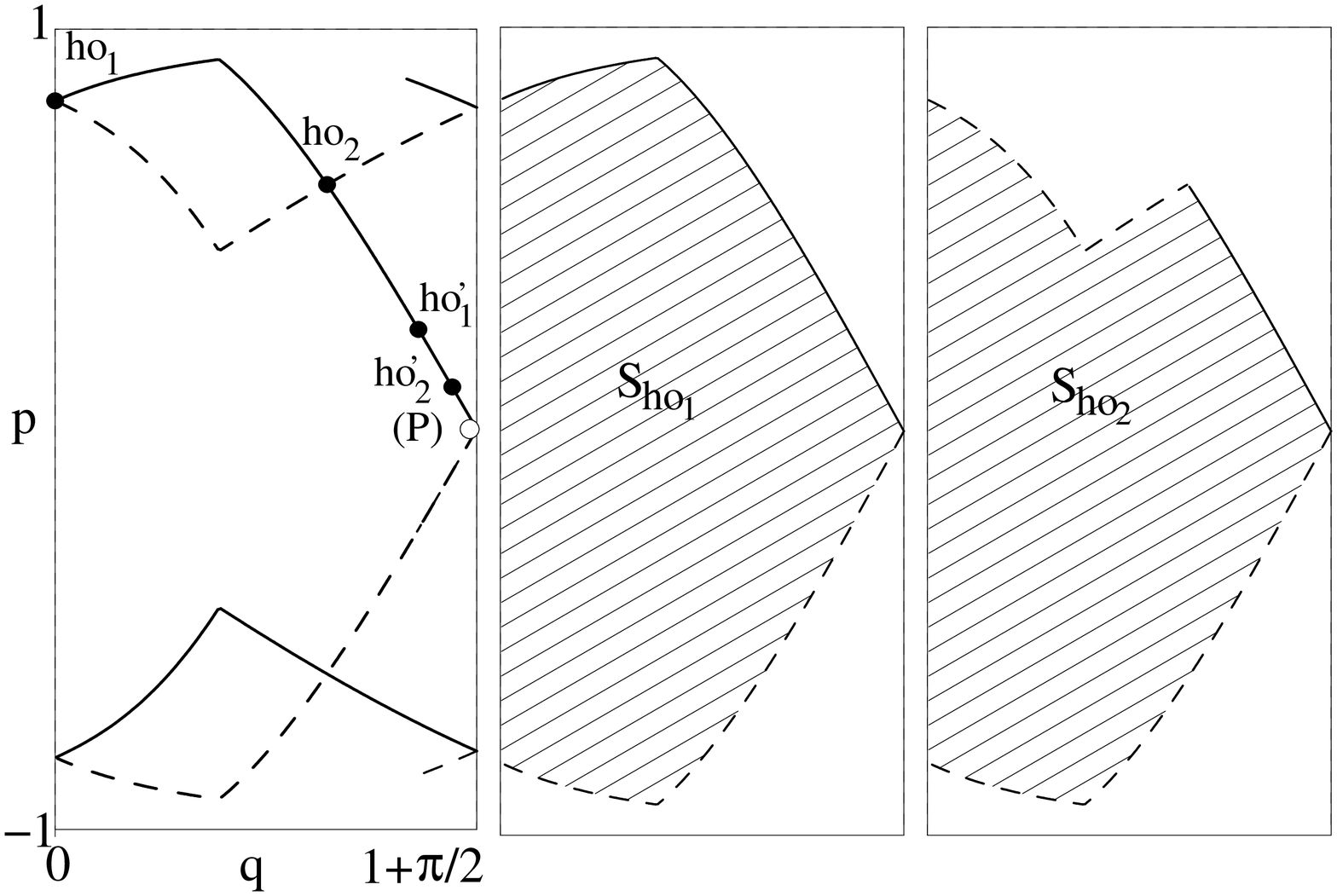} }
   \end{center}
 \caption{(Left) Detail of the phase space portrait showing the relevant
  elements related with the dynamics of the horizontal periodic orbit
  shown in Fig.~\protect\ref{fig:1}. \protect \\
  Label (P) indicates the position of the corresponding fixed point.
  The associated unstable and stable manifolds are represented
  in full and dashed line, respectively.
  They cross, with different topology, at the primary homoclinic
  points ${\rm ho}_1$ and ${\rm ho}_2$, which map into
  ${\rm ho}'_1$, ${\rm ho}''_1$, $\ldots$,
  and ${\rm ho}'_2$, ${\rm ho}''_2$, $\ldots$, respectively. \protect \\
  (Center and right) Primary homoclinic areas $S_{{\rm ho}_1}$ and
  $S_{{\rm ho}_2}$ associated to the homoclinic circuit for the horizontal
  periodic orbit.}
 \label{fig:2}
\end{figure}

A suitable phase space picture of the associated dynamics can be obtained
by defining a Poincar\'e surface of section consisting on plotting
the coordinates $(q,p)$ when the trajectory hits the boundary
(axis excluded to reconcile with the dynamics in the full version of the
stadium).
This is presented in Fig.~\ref{fig:2}, where some elements relevant to our
work are shown.
For example, the horizontal PO renders a fixed point located at
$(q,p)=(1+\pi/2,0)$, that appears labelled (P) in the plot.
%
From this point two manifolds, one unstable and the other stable, emanate.
They are plotted, with full and dashed line respectively, in the figure.
As can be seen, these two manifolds first cross at the two primary homoclinic
points, labelled ${\rm ho}_1$ and ${\rm ho}_2$ respectively in the figure,
and then at ${\rm ho}_1, {\rm ho}'_1, \ldots$ and
${\rm ho}_2, {\rm ho}'_2, \ldots$ (and their reflections with respect to
the line $p=0$).
These crossings form lobes (the first of which is the only one shown in
the figure) that by subsequent propagations with the Poincar\'e map form
the well known homoclinic tangle.
This tangle organizes all the dynamical complexity of the
system at the classical level, as shown in the early work of Poincar\'e.
These two infinite sequences of crossing points define the so--called primary
homoclinic orbits, two in this case.
As can be seen, they define in phase space two areas,
$S_{{\rm ho}_1}$ and $S_{{\rm ho}_2}$, which are shown as shaded regions
in the central and rightmost panels of Fig.~\ref{fig:2}, respectively.

\begin{figure}[b]
  \begin{center}
   \resizebox{0.95\columnwidth}{!}{ \includegraphics{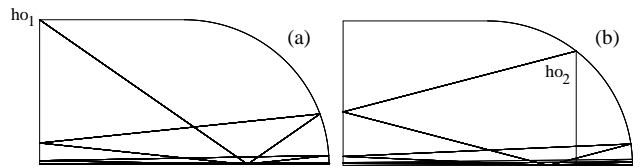} }
  \end{center}
 \caption{Primary homoclinic orbits passing respectively by ${\rm ho}_1$ (a),
  and ${\rm ho}_2$ (b).
  The position in configuration space of these two phase space points has
  also been indicated in the plots.}
 \label{fig:3}
\end{figure}

It is also interesting to examine these orbits in configuration space.
Since these orbits corresponds to infinitely long trajectories,
accumulating at both ends in the horizontal PO,
a complete calculation of such orbits it is not feasible.
Accordingly, we will present here only a shortened version of them that
captures, however, the main dynamical part of it.
These surrogates of the full orbits are computed in our case in the
following way.
We start by locating very accurately the position in phase space of
the first two crossings of the unstable and stable manifolds emanating
from the fixed point (P); this giving the position of the two
primary homoclinic points ${\rm ho}_1$ and ${\rm ho}_2$.
A piece of the required orbits is then calculated by propagating forward
in time the corresponding trajectories until they cross several more
times (6 in our case) the SOS.
The corresponding results are shown in Fig.~\ref{fig:3}.
Obviously, longer versions of the whole orbits can be produced
by simply enlarging the above described procedure.
This would involve taking the initial point closer to the fixed point
and ending the trajectory at a later time, so that more points on the
SOS at both ends of our previous approximation to the full orbit
are collected.
However, this would only complicate visually our figures with more
cumbersome orbits, adding very little to the dynamics that we are
now analyzing.
%
\section{Quantum mechanical aspects}
   \label{sec:quantum}

All quantum calculations performed in this work are carried out by imposing
Dirichlet conditions on the stadium boundary and Neumman conditions on the
(horizontal and vertical) axes.

To perform our study we first need a method (the tool) to systematically
construct a series of scar wave functions highly localized on a given PO.
Let us remark in this respect that the appearance of scarred eigenfunctions
becomes less frequent as the energy increases, being vanishingly small in
the semiclassical limit.
Accordingly, the scar wave functions that we need are, in general,
non--stationary and can be obtained, for example, as linear combination of
the eigenstates of the system, as described by Bogomolny \cite{Bogomolny}.

Moreover, very sophisticated methods for performing this task has also
been reported in the literature \cite{Polavieja,Vergini,PRL2}.
In this work we use the method developed in Ref.~\cite{PRL2},
which can be summarized as follows.
We start from ``tube wave functions'', $|\phi_{\rm tube} \rangle$,
covering the configuration region around the PO.
They are obtained from the semiclassical theory of resonances developed by
Vergini and Carlo \cite{Vergini,PRE}, and are calculated at the BS quantized
energies [see eq.~(\ref{eq:2})], which for our model can be expressed,
in terms of the wave numbers, as
%
\begin{equation}
  k_{\rm BS} = \frac{2\pi}{L_H} \left(n_H+\frac{\nu_H}{4} \right)\;,
  \label{eq:3}
\end{equation}
with $L_H=4$ (length of the horizontal PO) and $\nu=3$.
After that, we define a more elaborated scar function by incorporating
the associated averaged \cite{Polavieja} dynamical information up to
a given time, $T$, and in the energy window $E_{\rm BS}\pm \hbar/T$.
This is done by using the expression
%
\begin{equation}
  |\phi_{\rm scar} \rangle = \int_{-T}^{T}\; dt \;
     \cos \left(\frac{\pi t}{2T} \right) \;
     e^{i(E_{\rm BS}-\hat{H})t/\hbar} \;
     |\phi_{\rm tube} \rangle,
  \label{eq:4}
\end{equation}
where the cosine filter is introduced to minimize the energy dispersion
of the resulting functions.

To monitor the different properties and characteristics of
our scar functions, specially those aspects concerning the dynamical
information carried by them, we resort to quantum surfaces of section
(QSOS) obtained from suitable (quasi)probability density distributions,
in our case the Husimi function \cite{Husimi}.
Phase space representations for the wave functions described in this
section can be defined in a number of ways.
In this paper we will follow the procedure described in Ref.~\cite{tua},
which relies on the use of normal derivatives of the wave functions
evaluated at specific points on the boundary of the billiard, $(q,p)$.
The coherent states necessary to construct such representation (Husimi
functions) are then defined as:
%
\begin{equation}
  G_{q,p}(\bar{q}) = \left( \frac{1}{\pi\sigma^2} \right)^{1/4}
  \exp \left[-\frac{1}{2\sigma^2} (\bar{q}-q)^2+ip(\bar{q}-q)\right].
 \label{eq:5}
\end{equation}
This expression corresponds to a boundary wave packet centered at the point
($q,p$) on the surface of section, and represents a bounce off a given
boundary point with a specific tangential momentum.
Then, for a wave function with normal derivative on the boundary $\phi(q)$
(extended periodically to the real line), the corresponding Husimi function
is given by
%
\begin{equation}
  H(q,p) = \left| \langle G_{q,p}|\phi_{\rm scar} \rangle\right|^2
 \label{eq:6}
\end{equation}

%
\begin{figure}
  \begin{center}
   \resizebox{0.95\columnwidth}{!}{ \includegraphics{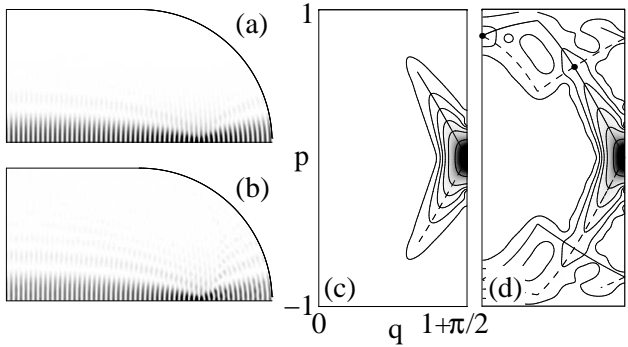} }
  \end{center}
 \caption{Scar function localized along the horizontal periodic orbit
   corresponding to the quantum number $n_H=54$, in configuration
   (a)--(b) and phase space (c)--(d).
   It has been computed with the aid of Eq.~(\protect\ref{eq:2}) using
   two different values of the time parameter $T$: (a)--(c) $t_E$,
   and (c) $1.5 t_E$, being $t_E$ the Ehrenfest time \protect\cite{tE}.}
 \label{fig:4}
\end{figure}

As an example, we present in Fig.~\ref{fig:4}, both in configuration
(a)--(b) and phase space (c)--(d), the scar wave function associated
to the wave number $k=86.01$,
value that been obtained using the BS quantization condition (\ref{eq:3})
with $n_H=54$.
Two values of the cut--off parameter, $T$, appearing in Eq.~(\ref{eq:4}),
have been used, being those equal to $T=t_E$ for (a)--(c) and to $1.5 t_E$
for (b)--(d), respectively.
As can be seen in parts (a) and (b), the scar wave functions appear
highly localized along the horizontal PO.
Moreover, it should be remarked that the ``quality'' of these wave functions
is very high since, for example, the focal point \cite{Heller2}
existing on the horizontal axis at approximately 3/4 of the PO total
length is clearly visible in both plots.
The reason for this has been thoroughly discussed in Ref.~\cite{Vergini},
and comes from the fact that the dynamics in the vicinity of the
scarring PO has been included in eq.~(\ref{eq:4}).
This is also illustrated in the results on the other two panels, (c)--(d),
of Fig.~\ref{fig:4}.
As can be seen there, the corresponding Husimi based QSOS is not only
localized on the fixed point (P), corresponding to the horizontal PO,
but also spreads in a substantial way along the paths of the associated
manifolds.
Moreover, this function spreads further along these manifolds as the time
parameter $T$ is increased.
Actually, for the lowest value considered here, $T=t_E$, the probability
distribution only covers the linear part of the manifolds, while for the
larger value $T=1.5t_E$ it extends much further, getting even to the
region where the primary homoclinic points, ${\rm ho}_1$ and ${\rm ho}_2$,
are located.

Let us now return to the main point addressed in this paper,
namely, what it is the fate of the probability density which after
being launched from a point in the vicinity of the horizontal PO,
instead of returning next to it in the first recurrences,
it is thrown out of this region.
As will be shown, this can be unveiled by analyzing the behavior of the
probability density for different values of the quantum number $n_H$,
or equivalently of the corresponding BS quantized wave number, $k$.
Instead of considering the whole distribution, we will simplify our
calculation by focussing the attention into a single relevant point.
Fixed point (P) seems to be the most representative one,
since it is in its vicinity that the density returning by the
recurrent mechanism discussed by Heller in his 1984 paper \cite{Heller1},
is coherently collected.
Let us remark once again that this is so because in the way we perform
our calculations, the action along the PO circuit is quantized.
%
\begin{figure}
  \begin{center}
   \resizebox{0.95\columnwidth}{!}{ \includegraphics[angle=-90]{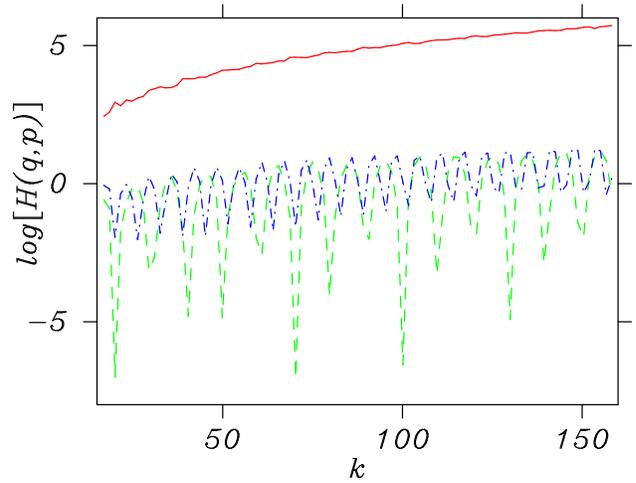} }
  \end{center}
 \caption{(color online)
   Logarithm of the amplitude of the Husimi based quantum surface of
   section for the scar wave functions evaluated at points
   P (red full line), ${\rm ho}_1$ (green dashed line) and ${\rm ho}_2$
   (blue dashed--dot line), as a function of the corresponding wave number,
   $k$.}
 \label{fig:5}
\end{figure}

\begin{figure}
  \begin{center}
   \resizebox{0.95\columnwidth}{!}{ \includegraphics[angle=-90]{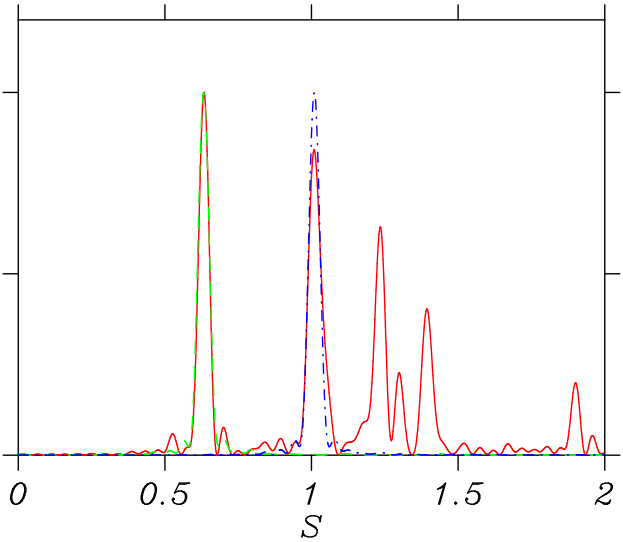} }
  \end{center}
 \caption{(color online)
   Fourier transform of the amplitudes in phase space presented in
   Fig.~\ref{fig:5} after having removed the slowly varying contribution
   as described in the text.}
 \label{fig:6}
\end{figure}

The results for 91 scar wave functions corresponding to $n=10-100$
($k=16.88-158.25$) are shown in red full line in Fig.~\ref{fig:5}.
In it we see that, superimposed to a slightly increasing overall
tendency, there is a small oscillatory behavior.
Although this is not well visible in the scale of the figure,
it is obvious when one considers its Fourier transform,
which is shown in Fig.~\ref{fig:6}.
At this respect, notice that the conjugate variable used, $S$,
has units of length, and plays the role of an action in our
calculations due to the unit system that we have chosen.
%

As can be seen, several peaks contribute to the oscillatory behavior of
$\log H$ computed at phase space point (P), being those at $S=0.633$ and
1.007 the most important ones.
The origin of this regularity can be now quite clearly understood.
In the first place, the overall value in the QSOS function, $H$, comes
from the scarring mechanism by recurrences along the (horizontal in this
case) PO, as described by Heller in Ref.~\cite{Heller1}.
The actual value may change slightly with $n_H$, but this variation is small,
since the essential constructive mechanism induced by the PO recurrences
is the same in all cases.
However, and as proposed by us, we have a second, superimposed mechanism.
This consists in recurrences due to longer, and increasingly more
complicated paths in phase space.
From all of them, the shortest possible are those taking place along the
homoclinic circuits defined by the two primary homoclinic orbits described
in Sect.~\ref{sec:classical}.
This second interference mechanism is controlled by the value of a different
action, and then the corresponding quantization condition does not coincide,
in general, with that of the scarring PO.
However, every so often the two quantization conditions will take place at
very close energies; this will correspond to a maximum in the value of
the Husimi function, $H$, computed at the phase point (P).
Obviously, at the midpoint between two consecutive quantized energies of
these, we will be in an ``antiquantization'' point, where the interference
of the quantum density when arriving at (P), after a recurrence along the
homoclinic circuit, will be destructive, and then we will find a minimum
in $H$.
Intermediate situation will then be found at the energies in the range
defined by the two quantized energies, and this is the origin of the
oscillatory behavior we are observing in the red full line of
Fig.~\ref{fig:5}.

To confirm numerically this interpretation, and explain why we are
observing two principal frequencies instead of one, let us consider
now the quantization condition(s) of the homoclinic circuit defined
by the manifolds emanating from (P) (see two rightmost panels in
Fig.~\ref{fig:2}).
As mentioned before the homoclinic orbits defining these circuits are
infinite in length, something which makes difficult to derive the
quantization condition we are looking for.
The way to deal with this problem was discussed in detail in Ref.~\cite{PRL1},
and it is based on an idea of Ozorio de Almeida \cite{Ozorio}, when
he tried to quantized the so--called ``homoclinic torus''.
Basically, it consists of taking advantage of the fact that the homoclinic
orbits both leave and goes back asymptotically to the horizontal orbit (P).
In this way, instead of calculating the action along the homoclinic orbits,
one can just compute the difference with the horizontal one, which is
already quantized.
Accordingly (see full details in \cite{PRL1})
\begin{equation}
    k S_{{\rm ho}_i} - \frac{\pi}{2} \nu_{{\rm ho}_i} = 2 \pi n,
       \qquad i=1,2,
 \label{eq:7}
\end{equation}
which is valid for both primary homoclinic orbits/circuits, ${\rm ho}_1$
and ${\rm ho}_2$.
In our case it is not difficult to show that $\nu_{{\rm ho}_1}=-1$ and
$\nu_{{\rm ho}_2}=0$ \cite{PRL1},
and the actual values of $S_{{\rm ho}_1}$ and $S_{{\rm ho}_2}$ can be
numerically calculated with the aid of Fig.~\ref{fig:2},
resulting equal to $-3.368\;390\;45$ and $-2.991\;142\;22$, respectively.
These two quantities are in excellent agreement with the two values,
0.633 and 1.007, found in the Fourier transform of $H$ evaluated at (P)
(see Fig.~\ref{fig:6}), after subtracting the length of the horizontal
orbit, $L_H=4$.

To conclude this section it should be pointed out that similar results
are obtained if apply the same type of analysis presented above
to the fluctuations of the Husimi function evaluated at
the two primary homoclinic points, ${\rm ho}_1$ and ${\rm ho}_2$.
The results are shown in Fig.~\ref{fig:5} with green dashed and
blue dashed--dot lines, respectively.
In these two cases, the oscillations are much more pronounced, but the Fourier
transform (Fig.~\ref{fig:6}) shows identical results.
Namely, for the first curve a single big peak, corresponding to
$S=0.633$ exists, while the oscillations for the second one are fully
accounted for with a single frequency at 1.007.
Both results are in perfect agreement with the results and conclusions
discussed before.
\section{Conclusions}
   \label{sec:conclusions}

In this paper, we have revisited the scarring mechanism proposed by Heller
in his 1984 paper \cite{Heller1}, in which he analyzed the role played
by recurrences along classical unstable POs.
We perform our study in the energy rather than in the time domain,
as it is customary when arguments based on the self--correlation function,
$C(t)$, are used.
For this purpose, sophisticated scarred wave functions, highly localized
along POs, have been used as a tool for our analysis.
Special attention is paid to the fate of the quantum probability density
first expelled from the unstable vicinity of the scarring PO,
showing the relevance of the associated homoclinic circuits for the
coherent, i.e.\ causing constructive interference, transport
of probability density in classically chaotic systems.
\begin{acknowledgement}
Support from MEC--Spain (under contracts MTM2006--15533 and
i--MATH CSD2006--32),
Comunidad de Madrid--Spain (SIMUMAT S-0505/ESP-0158),
Agencia Espa\~nola de Cooperaci\'on Internacional
(PCI--A/6072/06 and PCI--A/4872/06),
UAM--Grupo Santander,
CONICET--Argentina and UBACYT (X248) is fully acknowledged.
\end{acknowledgement}


\begin{thebibliography}{99}
\bibitem{tE} The Ehrenfest time is defined as
  $T_E=(2\lambda)^{-1}\ln (S/\hbar)$,
  $S$ being a typical action at the energy that is considered.
  This parameter constitutes a rough indicator of the amount of
  time needed by a wave packet to disperse over all the available
  configuration space range.
\end{thebibliography}
\end{document}